\documentclass{XrU2005}

\title{The XMM-Newton Slew Survey: towards the XMMSL1 catalogue}
\author[1]{M.P. Esquej}
\author[1]{B. Altieri}
\author[1]{D. Bermejo}
\author[2]{M.J. Freyberg}
\author[1]{V. Lazaro}
\author[3]{A.M. Read}
\author[1]{R.D. Saxton}
\affil[1]{European Space Agency (ESA), European Space Astronomy Centre, Villafranca, Apartado 50727,  28080 Madrid, Spain}

\affil[2]{Max-Planck-Institut f\"{u}r extraterrestrische Physik, Giessenbachstrasse 1, 85784 Garching, Germany}
\affil[3]{Dept. of Physics and Astronomy, Leicester University, Leicester LE1 7RH, U.K.}

\begin{document}


\maketitle

\begin{abstract}
The XMM-Newton satellite is the most sensitive X-ray observatory flown to date 
due to the great collecting area of its mirrors coupled with the high quantum 
efficiency of the EPIC detectors. It performs slewing manoeuvers between observation 
targets tracking almost circular orbits through the ecliptic poles due to the Sun constraint.
Slews are made with the EPIC cameras open and the other instruments closed,
operating with the observing mode set to the one of the previous pointed observation and the medium 
filter in place.

Slew observations from the EPIC-pn camera in FF, eFF and LW modes provide data, resulting 
in a maximum of 15 seconds of on-source time. These data can be used to give a uniform survey of the X-ray sky,
at great sensitivity in the hard band compared with other X-ray all-sky surveys.

\end{abstract}

\section{Introduction}
XMM-Newton traces slewing paths over the sky while manoeuvering with both EPIC-pn 
and EPIC-MOS cameras open. 
Data from slew observations are recorded into Slew Data Files (SDF), which have been stored 
in the XMM-Newton Science Archive (XSA) from revolution 314.
Not all these data are scientifically useful and 
data from the EPIC-MOS cameras are now used for calibration purposes.

This paper describes the EPIC-pn slew data processing strategy, used to give a uniform coverage over the sky, 
in order to create the first catalogue of slew detections with XMM-Newton \citep{Freyberg}.
It also reports on the current status and
scientific utility of the survey.

\section{Observations and data analysis}

The optimum source searching strategy derived for
slew data processing \citep{Read,Saxton} is described below. The 
attitude reconstruction
and spurious detections are dealt with in the corresponding subsections.

Data from the EPIC-pn camera are only used due to the faster 
readout in its observing modes and its high effective area with respect to the EPIC-MOS cameras.
In particular, only FF, eFF and LW modes are used because the other EPIC-pn modes
are not appropiate for source determination.
The characteristic low background of the observations (average ~0.1 cts/arcmin$^{2}$)
and the tight PSF of the telescopes provide good sensitivity to detect extended sources \citep{Lazaro} .
Nevertheless, slews performed at times of enhanced solar activity have been rejected 
in the current processing although they 
are hoped to be included in the future. 
Slew observations are divided into $\sim$~1 square degree event files before processing in order to
get accurate positions over the sky. 
A near standard pipeline eboxdetect/emldetect tuned for $\sim$~zero background was performed on images
containing only single events (pattern 0) in the 0.2-0.5 keV energy range and single plus double events 
(patterns 0$-$4) in the 0.5-12 keV band.
Three 
different energy bands are source searched independently: total band (0.2$-$0.5 keV), soft band (0.2$-$2 keV) and
hard band (2$-$12 keV).

\subsection{Attitude reconstruction}
The attitude reconstruction is crucial in the determination of source coordinates.
After further investigation we concluded that during slews an
attitude reconstruction slightly different than for pointed observations
had to be performed.
The optimal attitude file for reconstructing the astrometry in slew observations is 
the Raw Attitude File (RAF) with 0.75 seconds subtracted from every entry, a timing error 
that is due to a delay of the star tracker CCDs.

\begin{figure}[ht]
\centering
\resizebox{\hsize}{!}{\rotatebox[]{270}{\includegraphics[clip=true]{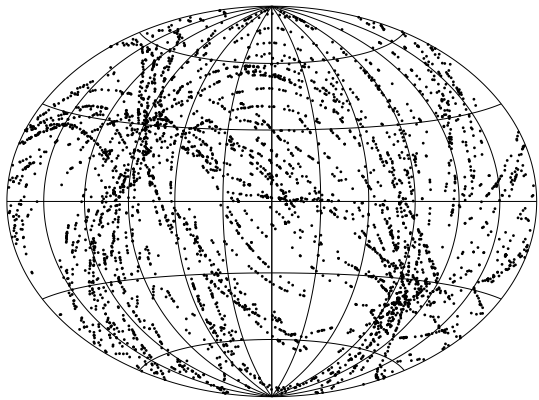}}}
\caption{Aitoff projection of the distribution of all XMMSL1 detections in the total band.
	}
\label{aitoff}
\end{figure}

\subsection{Spurious detections}
Systematic effects in the instrument and detection software lead to a number of spurious detections
that are outlined below.
In the current slew pipeline unreal sources due to optical loading and detector flashes
are directly rejected during processing by using only single-pixel events (pattern 0) below 0.5 keV.

False detection: detections not verified through visual
inspection. Within bright and/or extended source: multiple detections of the same object. 
Position suspect: sources located at the edge of an image and others. Background related:
sources positioned in localised flared images.

\section{The XMM-Newton Slew catalogue}
Images and exposure maps have been source searched for 219 slew observations producing
4179 detections in the total band (Fig.~\ref{aitoff}), 2750 in the soft band 
and 844 in the hard band. The number of real sources is under investigation as spurious detections
are currently being flagged.
The sky coverage is $\sim$~6300 square degrees which means $\sim$~15\% of 
the whole sky, indicating a source density of about 0.65 sources per square degree.

In order to check the quality of our detections
we correlated the 2178 non-extended sources with det\_ml$>$10 (sigma $~$3.9) with different catalogues.
It was found that $\sim$~56\% of the sources have a RASS counterpart
within 60 arcsec, with 68\% of matches lying within 15 arcsec. Furthermore, correlations with the
astronomical database SIMBAD show that 68\% of the matches lie within 8 arcsec.
These correlations also indicate a great variety of detected objects 
during slews, including AGN, galaxies, cluster of galaxies, LMXB and SNR among others.

\begin{figure}
\centering
\resizebox{\hsize}{!}{\rotatebox[]{270}{\includegraphics[clip=true]{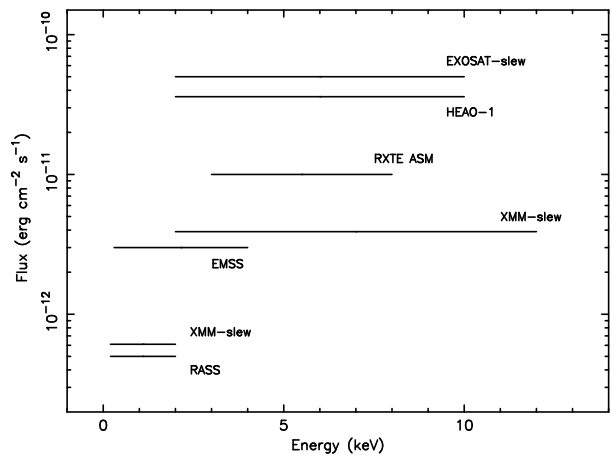}}}
\caption{Flux limits of the X-ray large area surveys. Fluxes for the XMM-slew survey 
	have been calculated for a source with det\_ml=10 and passing through the centre of the 
	field of view. These fluxes were derived from count rates based on energy conversion factors 
	assuming an absorbed power-law model with $N_H$ = $3.0 \times10^{20}$ $cm^{2}$ and slope 1.7}
\label{flux}
\end{figure}

The sensitivity of the survey in the different bands was obtained and flux limits at det\_ml of 10(8)
were compared with those of other X-ray all-sky surveys (Fig. ~\ref{flux}). The soft X-ray band detection limit is
$6(4.5)\times10^{-13}$ $erg$ $s^{-1}$ $cm^{-2}$, comparable to the one of the ROSAT bright source catalogue \citep{Voges}.
The sensitivity of slew detections is particularly evident for the hard X-ray band whose limit
is the deepest ever $4(3)\times10^{-12}$ $erg$ $s^{-1}$ $cm^{-2}$ (ten times deeper than EXOSAT, HEAO-1).
	
All detected sources will comprise the first XMM-Newton catalogue
derived from slew observations, the XMM-Newton Slew 1 (XMMSL1). It is expected to be published by the end of
2005 and updated when more slews are available to finally have an all-sky survey.

\end{document}